\documentclass[12pt]{article}

\usepackage{amsmath,amssymb}
\newcommand{\mathscr}{\cal}

\begin{document}

\renewcommand{\square}{\vrule height 1.5ex width 1.2ex depth -.1ex }


\newcommand{\II}{\leavevmode\hbox{\rm{\small1\kern-3.8pt\normalsize1}}}  

\newcommand{\CC}{{\mathbb C}}
\newcommand{\RR}{{\mathbb R}}
\newcommand{\NN}{{\mathbb N}}
\newcommand{\QQ}{{\mathbb Q}}
\newcommand{\ZZ}{{\mathbb Z}}

\newcommand{\Af}{{\mathfrak A}}
\newcommand{\Tf}{{\mathfrak T}}


\newcommand{\CoinfM}{C_0^\infty(M)}
\newcommand{\CoinfN}{C_0^\infty(N)}
\newcommand{\Coinfd}{C_0^\infty(\RR^d\backslash\{ 0\})}
\newcommand{\Coinf}[1]{C_0^\infty(\RR^{#1}\backslash\{ 0\})}
\newcommand{\CoinX}[1]{C_0^\infty({#1})}
\newcommand{\Coin}{C_0^\infty(0,\infty)}


\newtheorem{Thm}{Theorem}[section]
\newtheorem{Def}[Thm]{Definition}
\newtheorem{Lem}[Thm]{Lemma}
\newtheorem{Prop}[Thm]{Proposition}
\newtheorem{Cor}[Thm]{Corollary}


\renewcommand{\theequation}{\thesection.\arabic{equation}}
\newcommand{\sect}[1]{\section{#1}\setcounter{equation}{0}}


\newcommand{\AAA}{{\cal A}}
\newcommand{\AAt}{\widetilde{\cal A}}

\newcommand{\BB}{{\cal B}}
\newcommand{\CCC}{{\cal C}}

\newcommand{\DD}{{\mathscr D}}
\newcommand{\EE}{{\mathscr E}}

\newcommand{\FF}{{\cal F}}
\newcommand{\GG}{{\cal G}}
\newcommand{\KK}{{\cal K}}
\newcommand{\OO}{{\cal O}}
\newcommand{\MM}{{\cal M}}
\newcommand{\HH}{{\cal H}}
\newcommand{\PP}{{\cal P}}
\newcommand{\VV}{{\cal V}}
\newcommand{\WW}{{\cal W}}


\newcommand{\kvec}{{\bf k}}
\newcommand{\xvec}{{\bf x}}
\newcommand{\rvec}{{\bf \hat{r}}}
\newcommand{\nvec}{{\bf n}}
\newcommand{\kkvec}{\mbox{\boldmath $\kappa$}}

\newcommand{\eb}{{\boldsymbol{e}}}
\newcommand{\gb}{{\boldsymbol{g}}}
\newcommand{\nb}{{\boldsymbol{n}}}
\newcommand{\tb}{{\boldsymbol{t}}}
\newcommand{\vb}{{\boldsymbol{v}}}
\newcommand{\zb}{{\boldsymbol{z}}}
\newcommand{\etb}{{\boldsymbol{\eta}}}
\newcommand{\nub}{{\boldsymbol{\nu}}}


\newcommand{\Pt}{\hbox{\sf P}}
\newcommand{\Tt}{\hbox{\sf T}}
\newcommand{\SSS}{{\sf S}}


\newcommand{\Dal}{\fbox{\phantom{${\scriptstyle *}$}}}
\newcommand{\vth}{\vartheta}
\newcommand{\vep}{\varepsilon}
\newcommand{\Ran}{{\rm Ran}\,}
\newcommand{\supp}{{\rm supp}\,}
\newcommand{\Ann}{{\rm Ann}\,}

\newcommand{\Cgt}{C_{\gb}^{T+}}
\newcommand{\Cgs}{C_{\gb}^{S+}}
\newcommand{\Cet}{C_{\etb}^{T+}}
\newcommand{\Ces}{C_{\etb}^{S+}}

\newcommand{\stack}[2]{\substack{#1 \\ #2}}

\newcommand{\wcheck}[1]{\stackrel{\smile}{#1}}

\begin{titlepage}
\renewcommand{\thefootnote}{\fnsymbol{footnote}}

\vspace{0.1in}
\LARGE
\center{Bisolutions to the Klein--Gordon Equation \\ and Quantum Field 
Theory on 2-dimensional Cylinder Spacetimes} 
\Large

\vspace{0.2in}
\center{C.J. Fewster\footnote{E-mail: {\tt cjf3@york.ac.uk}}}
\center{Department of Mathematics, University of York, \\
Heslington, York YO1 5DD, United Kingdom.}

\normalsize
\center{April 3, 1998}

\begin{abstract}
We consider 2-dimensional cylinder spacetimes whose metrics differ from the flat
Minkowskian metric within a compact region $K$. By choice of time orientation,
these spacetimes may be regarded as either globally hyperbolic {\em timelike
cylinders} or nonglobally hyperbolic {\em spacelike cylinders}. For generic
metrics in our class, we classify all possible candidate quantum field algebras
for massive Klein--Gordon theory which obey the F-locality condition introduced
by Kay. This condition requires each point of spacetime to have an 
intrinsically globally hyperbolic neighbourhood $N$ such that the commutator
(in the candidate algebra) of fields smeared with test functions supported in
$N$ agrees with the value obtained in the usual construction of Klein--Gordon
theory on $N$. 

By considering bisolutions to the Klein--Gordon equation, we prove that 
generic timelike cylinders admit a unique F-local algebra -- namely the algebra
obtained by the usual construction -- and that generic spacelike cylinders do
not admit any F-local algebras, and are therefore non F-quantum compatible.
Refined versions of our results are obtained for subclasses of metrics
invariant under a symmetry group. Thus F-local field theory on 2-dimensional
cylinder spacetimes essentially coincides with the usual globally hyperbolic
theory. In particular the result of the author and Higuchi that the Minkowskian
spacelike cylinder admits infinitely many F-local algebras is now seen to
represent an anomalous case.

\end{abstract}

\end{titlepage}

\sect{Introduction}

During the last few years there has been much interest in classical and
quantum dynamics on nonglobally hyperbolic spacetimes, with attention
focussing on those spacetimes which contain closed timelike curves and
which therefore provide mathematical models for the concept of a `time
machine'. The usual formulation of quantum field theory breaks down on
these spacetimes owing to the absence of global advanced-minus-retarded
fundamental solutions to the corresponding classical field equations, so
any study of quantum theory on nonglobally hyperbolic spacetimes
must be undertaken within a generalisation of the usual theory. In this
paper, we will study quantum field theory on both globally hyperbolic
and nonglobally hyperbolic 2-dimensional cylinder manifolds within 
a framework for quantum field 
theory on not-necessarily globally hyperbolic spacetimes proposed by
Kay~\cite{Kay}. 
(See~\cite{Yurt,FPS,HawkSS,FHW} and references therein
for other approaches to the formalism and interpretation of quantum
theory in the presence of chronology violation.) 

To be specific, consider the linear covariant Klein--Gordon equation
\begin{equation}
(\Dal_\gb + m^2)\phi = 0 
\label{eq:introKG}
\end{equation}
on a Lorentzian spacetime $(M,\gb)$. If $(M,\gb)$ is globally
hyperbolic, there exists a (distributional) global
advanced-minus-retarded fundamental solution $\Delta_\gb(x,x')$
to~(\ref{eq:introKG})~\cite{YCB,Dim} which plays a key r\^ole in the
quantisation of real scalar field theory on $(M,\gb)$ via the
commutation relation 
\begin{equation}
[\phi(f_1),\phi(f_2)]=i\Delta_\gb(f_1,f_2)\II
\label{eq:comm}
\end{equation}
obeyed by smeared fields. Kay's proposal for field theory on more
general spacetimes is that~(\ref{eq:comm}) should be replaced by a local
equivalent in the following way. A candidate quantum field algebra
for Eq.~(\ref{eq:introKG}) is said to be {\em F-local} if
each point in $M$ has an intrinsically globally hyperbolic
neighbourhood $N$ such that~(\ref{eq:comm}) holds for test functions
$f_i$ supported in $N$ and with $\Delta_\gb$ replaced by
$\Delta_{\gb|_N}$, the advanced-minus-retarded fundamental solution for
Eq.~(\ref{eq:introKG}) on the globally hyperbolic spacetime
$(N,\gb|_N)$. If $(M,\gb)$ admits an F-local algebra for
a given field theory it is said to be {\em F-quantum compatible} with that
theory. (It is possible for a spacetime to be F-quantum compatible with
massless Klein--Gordon theory but not with the massive theory~\cite{FHK}.) 
All globally hyperbolic spacetimes are trivially F-quantum
compatible with Klein--Gordon theory (for any mass) by virtue of the
usual construction. 

Kay argued in~\cite{Kay} that only an F-quantum compatible spacetime (or
one obeying a similar condition) could arise as a semi-classical
approximation to a state of quantum gravity. Thus the result proved
in~\cite{Kay} 
that 2-dimensional Misner space fails to be F-quantum compatible,
and the general 4-dimensional results proved by Kay, Radzikowski and
Wald~\cite{KRW} showing that no spacetime containing a 
compactly generated Cauchy horizon can be F-quantum compatible
may be taken as evidence that such spacetimes are unphysical. 
Further evidence for this view is provided by other results 
established in~\cite{KRW} which prove that, on a spacetime
with compactly generated Cauchy horizon, the stress energy tensor
for any quantum 2-point function which is well-behaved (i.e., Hadamard)
in the initial globally hyperbolic region must be ill-defined at
certain points of the Cauchy horizon (see also~\cite{CraKa1,CraKa2} for
examples of this behaviour). 

These results provide strong evidence in support of Hawking's {\em
chronology protection conjecture}~\cite{HawkCP} that spacetimes 
with compactly generated Cauchy horizons (corresponding to the
notion of a `manufactured' time machine) are
unphysical. Nonetheless there are many nonglobally hyperbolic spacetimes  
not covered by these results. The simplest cases are the 2- and
4-dimensional Minkowskian {\em spacelike cylinders}~\cite{Kay}; namely,
2- or 4-dimensional Minkowski 
space quotiented by a timelike translation. These spacetimes do not
contain an initially globally hyperbolic region and therefore have no
Cauchy horizon, thus avoiding the no-go
theorems of~\cite{KRW}. In~\cite{Kay}, Kay used Huygens' principle
to show that the 4-dimensional spacelike cylinder is indeed
F-quantum compatible with {\em massless} Klein--Gordon theory. 
Subsequently, the present author and Higuchi~\cite{FH} used rather
different methods to establish F-quantum compatibility of the 
2- and 4-dimensional spacelike cylinders with massless and massive 
fields.\footnote{The 2-dimensional massless result was also known to Kay
using different methods.} Indeed, the construction used showed that
infinitely many F-local algebras exist on these spacetimes. 
These results demonstrated that the class of F-quantum compatible
spacetimes is strictly larger than the class of globally hyperbolic
spacetimes, but left open the question as to exactly how weak F-quantum  
compatibility is relative to global hyperbolicity.  

Some progress in answering this question has now been made by the
author, Higuchi and Kay, and will be described fully in~\cite{FHK}. It
appears that the F-quantum compatibility of the 2- and 4-dimensional
Minkowskian spacelike cylinders is unstable against certain classes of
metric perturbations. The present paper is an
outgrowth of~\cite{FHK} and consists of a full and rigorous treatment of
the massive Klein--Gordon equation on 2-dimensional cylinder spacetimes. 

To be specific, let the cylinder manifold $M$ be the quotient of $\RR^2$
with Cartesian coordinates $(t,z)$ by the translation $z\mapsto z+2\pi$.
We consider a class of smooth metrics on $M$ which agree with the
Minkowski metric
$\etb$ outside a compact region $K$ and are 
globally hyperbolic with respect to surfaces of constant $t$. Inside
$K$, the metrics may differ greatly from $\etb$. By choice of time
orientation, the resulting spacetimes are either globally hyperbolic
{\em timelike cylinders} (with $z$ interpreted as a spatial coordinate)
or nonglobally hyperbolic {\em spacelike cylinders} (with $z$
interpreted as a temporal coordinate). Our principal results are that
generic\footnote{We use this term in its
topological sense: a subset of a topological space is {\em generic} if
it contains a countable intersection of open dense sets (see,
e.g.,~\cite{GChoqu}).} timelike cylinders in our class admit a unique
F-local algebra --
the usual field algebra -- for massive Klein--Gordon theory, but that
generic spacelike cylinders are F-quantum incompatible with that theory.
Indeed, similar results hold for the weaker notion of a {\em locally
causal} algebra introduced in Section~\ref{sect:AQFT} which replaces
Eq.~(\ref{eq:comm}) with commutativity of fields at local spacelike
separation. We also refine our results to discuss cylinder spacetimes
with some degree of $z$-translational invariance. For generic timelike
cylinders in these classes we classify all possible F-local algebras in 
terms of distributions on the symmetry group; generic spacelike
cylinders still fail to be F-quantum compatible with massive
Klein--Gordon theory.  

These results show that F-locality does not provide significantly
more freedom than is afforded by the usual globally hyperbolic theory on 
these 2-dimensional cylinder manifolds, at least for massive
Klein--Gordon theory. It also appears that the  
F-quantum compatibility of the 2-dimensional Minkowskian spacelike
cylinder with massive Klein--Gordon theory~\cite{FH} is an anomalous
case, presumably due to its full spacetime translational invariance. We
have not found any other F-quantum compatible spacelike cylinder, and
conjecture that there are in fact no others. The situation for massless 
fields and for 4-dimensional cylinder spacetimes is discussed
in~\cite{FHK}. 

The structure of the paper is as follows. Section~\ref{sect:prelim}
contains some preliminaries relating to distributions on manifolds and
causal structure, while Section~\ref{sect:AQFT} briefly reviews the
algebraic approach to quantum field theory on curved spacetimes and
Kay's F-locality proposal~\cite{Kay}. In addition we introduce the
notion of a locally causal algebra. In Section~\ref{sect:TSC} we define
our class of timelike and spacelike cylinders and discuss various spaces
of weak (bi)solutions to the Klein--Gordon equation on these spacetimes,
and also develop a scattering formalism for this equation.
Sections~\ref{sect:lcbs} and~\ref{sect:results} form the main part of
the paper. For generic metrics in our classes, we use a unique
continuation result for Klein--Gordon
bisolutions~\cite{F} and the scattering formalism to classify all
bisolutions with the local causality property in
Section~\ref{sect:lcbs}. This leads into Section~\ref{sect:results}, in
which we reduce questions concerning F-local and locally causal algebras
to analogous questions concerning bisolutions with the corresponding
properties, enabling the classification of all F-local and locally causal
algebras on generic timelike and spacelike cylinders. We conclude with a
discussion of our results in Section~\ref{sect:concl}. There is one
Appendix, in which we establish a property of the scattering
operator employed in the main text. 

\sect{Preliminaries}\label{sect:prelim}

We begin by introducing the various spaces of test functions and 
distributions to be used in later sections. If $M$ is any paracompact 
$C^\infty$-manifold, $\EE^p_q(M)$ will denote the space of smooth complex
valued tensor fields of type $(p,q)$ on $M$, equipped with its usual
Fr\'echet space topology (see XVII.2 of Dieudonn\'e~\cite{Dieud3}). In
particular, we denote the space of scalar fields by $\EE(M)$.
Next, for any compact $K\subset M$, $\DD(K)$ is defined to be
the subspace of $\EE(M)$ consisting of functions vanishing outside
$K$. Each $\DD(K)$ is again a Fr\'echet space. Choosing an
increasing sequence $K_m$ of compact subsets of $M$ with $K_m
\subset {\rm int}\, K_{m+1}$ and $M=\bigcup_{m\ge 1} K_m$, we define
$\DD(M)$ to be the strict inductive limit of the spaces $\DD(K_m)$ 
(see, e.g., Sect.~V.4 in~\cite{RSi}). 
The construction is independent of the particular sequence
of compact sets used. 

We define the space of distributions\footnote{Sometimes (as in \S 6.3
of~\cite{Horm}) the dual of $\DD(M)$ is called the space of distribution 
densities, and the term distribution is used for the dual of the space
of smooth compactly supported densities.} $\DD'(M)$ to be the
topological dual of $\DD(M)$; elements of $\EE'(M)$, the topological
dual of $\EE(M)$, are precisely the distributions of compact support. 
We topologise $\DD'(M)$ and $\EE'(M)$ with their weak-$*$ topologies. 

Note that smooth functions on $M$ are not canonically embedded
in $\DD'(M)$. Instead, every positive density $\rho$ on $M$ defines
an embedding $\iota:C^\infty(M)\to\DD'(M)$ by
\begin{equation}
(\iota F)(f)=\int_M Ff\rho
\end{equation}
for all $f\in\DD(M)$. In particular, if $\gb$ is a metric on $M$, we will
write $\iota_\gb$ for the embedding corresponding to the density
$|\det \gb_{ab}|^{1/2}$.

A {\em bidistribution} on $M$ is a bilinear, separately continuous
functional $\Gamma(\cdot,\cdot)$ on $\DD(M)\times\DD(M)$. The space of
bidistributions is denoted $\DD^{(2)}{}'(M)$ and may be identified with
$(\DD(M)\otimes\DD(M))'$ when $\DD(M)\otimes\DD(M)$ is given the
projective tensor product topology (see \S III.6 of
Schaefer~\cite{Schaef}). We endow $\DD^{(2)}{}'(M)$ with the weak-$*$
topology arising from this identification, and also use the notations
$\Gamma(f,g)$ and $\Gamma(f\otimes g)$ interchangeably. 
By the kernel theorem (XXIII.9 in~\cite{Dieud7}) $\DD^{(2)}{}'(M)$ may
also be identified with $\DD'(M\times M)$. 

Next, we briefly review some notions of causal structure (see,
e.g.,~\cite{ONeill}). If $M$ is a 2-dimensional $C^\infty$-manifold, a
Lorentzian (that is, signature zero) metric $\gb$ on $M$
is said to be {\em time-orientable} if there exists a 
global continuous choice (called a choice of {\em time orientation}) 
$p\mapsto C_{\gb,p}^+\subset T_pM$ of closed convex cones obtained
at each $p\in M$ as the closure of a component of the set 
$\{\vb\in T_pM\mid \gb_p(\vb,\vb)\not=0\}$ in the tangent space
$T_pM$. In 2-dimensions there four possible
local choices of time orientation at each point. Given a global time
orientation $C_\gb^+$, a smooth curve in $M$ is said to be future
directed and causal (resp., timelike) if its tangent vector at each
point $p$ on the curve lies in the local causal cone $C_{\gb,p}^+$
(resp., in the interior of $C_{\gb,p}^+$). The causal future (past)
$J_\gb^+(A)$ ($J_\gb^-(A)$) of a subset $A\subset M$ with respect to
$C_\gb^+$ is the
union of $A$ with all points lying on future (past) directed causal
curves from $A$. A {\em Cauchy surface} in $M$ is a subset met exactly
once by every inextendible timelike curve in $M$. If $M$ admits a smooth 
foliation by $C^\infty$-Cauchy surfaces, the triple $(M,\gb,C_\gb^+)$ is 
said to be {\em globally hyperbolic}~\cite{Dieck}. A subset $N$ of a
time oriented spacetime $(M,\gb,C_\gb^+)$ is said to be {\em
intrinsically globally hyperbolic} if $(N,\gb|_N,C_\gb^+|_N)$ is
globally hyperbolic as a spacetime in its own right. 

The {\em Klein--Gordon operator} $P_\gb$ on $(M,\gb)$ is 
\begin{equation}
P_\gb=\Dal_\gb+\mu.
\end{equation}
Here $\mu\in\RR$, and the D'Alembertian $\Dal_\gb$ takes the form
\begin{equation}
\Dal_\gb = g^{-1/2} \partial_a g^{1/2}(\gb^{-1})^{ab}\partial_b  
\end{equation}
in local coordinates, where $g=-\det\gb_{ab}$ and $\gb^{-1}$
is the (unique) inverse to $\gb$, namely the
smooth tensor field of type $(2,0)$ such that $(\gb^{-1})^{ab}
\gb_{bc} = \delta^a{}_c$.\footnote{Normally
$(\gb^{-1})^{ab}$ is written as $\gb^{ab}$, but this notation is
inappropriate when more than one metric is under consideration,
as will be the case below.} The operator $P_\gb$ has the
self-adjointness property
\begin{equation}
(\iota_\gb f)(P_\gb h) = (\iota_\gb P_\gb f)(h)
\label{eq:saness}
\end{equation}
for $f,h\in\DD(M)$, where $\iota_\gb$ embeds 
$\EE(M)$ in $\DD'(M)$. If $(M,\gb,C_\gb^+)$ is globally
hyperbolic then~\cite{YCB,Dim} there exist continuous maps
$\Delta_\gb^\pm: \DD(M)\to\EE(M)$ (the retarded ($+$) and advanced ($-$)
Green functions for $P_\gb$) such that for all $f\in\DD(M)$,
\begin{equation}
P_\gb \Delta_\gb^\pm f= \Delta_\gb^\pm P_\gb f=  f
\end{equation}
and
\begin{equation}
\supp \Delta_\gb^\pm f\subset J_\gb^\pm(\supp f).
\end{equation}
Furthermore, $\iota_\gb\Delta_\gb^+f$ 
($\iota_\gb\Delta_\gb^- f$), is the unique element $\varphi$ of
$\DD'(M)$ having past (future) compact\footnote{A set $A$ is past
(future) compact with respect to $C_\gb^+$ if $J^-_\gb(\{p\})\cap A$
(resp., $J^+_\gb(\{p\})\cap A$) is 
compact for all $p\in M$.} support with respect to $C_\gb^+$ and obeying 
$\varphi(P_\gb h)= (\iota_\gb f)(h)$ for all $h\in\DD(M)$. 
The dual maps $\Delta_\gb^{\pm}{}':\EE'(M)\to\DD'(M)$ therefore obey
$\Delta_\gb^\pm{}'\iota_\gb = \iota_\gb\Delta_\gb^\mp$. 

The {\em advanced-minus-retarded fundamental solution} is
the map $\Delta_\gb:\DD(M)\to\EE(M)$ given by
$\Delta_\gb=\Delta_\gb^--\Delta_\gb^+$. This map defines a
bidistribution, also denoted $\Delta_\gb$, by
\begin{equation}
\Delta_\gb(f_1,f_2) = (\iota_\gb\Delta_\gb f_1)(f_2),
\end{equation}
which is antisymmetric in $f_1$ and $f_2$ owing to the formula
$\Delta_\gb'\iota_\gb =- \iota_\gb\Delta_\gb$ and is therefore a
bisolution for $P_\gb$, that is, $\Delta_\gb(P_\gb f_1,f_2)=
\Delta_\gb(f_1,P_\gb f_2) = 0$ for all $f_i\in\DD(M)$.
In addition, the support properties of
$\Delta_\gb^\pm$ force $\Delta_\gb(f_1,f_2)$ to vanish whenever
the support of $f_1$ is spacelike separated from that of $f_2$
with respect to $C_\gb^+$. 

\sect{Locally Causal and F-local Algebras}\label{sect:AQFT}

In this section, we give a brief review of algebraic quantum field
theory on globally hyperbolic curved spacetimes, and also describe Kay's 
F-locality condition. For a more detailed description see~\cite{Kay}. In
addition, we introduce a weakened version of F-locality which we call
`local causality'. 

Let $M$ be a (2-dimensional) paracompact $C^\infty$-manifold.
The test functions $\DD(M)$ label a set of abstract objects
$\{\phi(f)\mid f\in\DD(M)\}$ which will be interpreted as smeared
fields. The $\phi(f)$ are used to generate a free $*$-algebra $\Af(M)$
over $\CC$ and with $\II$, which is topologised in a natural way using
the topology of $\DD(M)$. In this topology, addition, multiplication,
conjugation are continuous operations in $\Af(M)$, and
$f\mapsto \phi(f)$ is continuous from $\DD(M)$ to $\Af(M)$. 
We will describe any quotient of $\Af(M)$ by a
closed (two-sided) $*$-ideal as a {\em $*$-algebra of smeared fields on
$M$}. Thus these algebras consist of
(congruence classes of) complex polynomials in the $\phi(f)$'s,
their conjugates $\phi(f)^*$ and the identity $\II$. The topology has
been chosen to ensure that if $\AAA$ is a $*$-algebra of smeared fields
then its topological dual $\AAA'$ separates the points of
$\AAA$.\footnote{That is, $a_1=a_2$ in $\AAA$ if and only if 
$\omega(a_1)=\omega(a_2)$ for all $\omega\in\AAA'$.} The
original development of F-locality~\cite{Kay} did not assume any
topology on field algebras; here it will play a useful r\^ole in linking 
the distributional and algebraic aspects of the argument. In addition,
smeared fields in~\cite{Kay,FH} were labelled by real-valued test
functions rather than complex-valued test functions as here. This
difference is in fact inessential for our purposes. 

Now suppose that a Lorentzian metric $\gb$ and time orientation 
$C_\gb^+$ are specified so that $(M,\gb,C_\gb^+)$ is globally 
hyperbolic. The {\em usual field algebra}, $\AAA(M,\gb)$, for real 
linear Klein--Gordon quantum field theory on $(M,\gb,C_\gb^+)$ is
defined to be the quotient of $\Af(M)$ by 
the following relations (which generate a closed $*$-ideal in $\Af(M)$):
\begin{list}{(Q\arabic{enumii})}{\usecounter{enumii}}
\item Hermiticity: $(\phi(f))^*=\phi(\overline{f})$ for all $f\in\DD(M)$
\item Linearity: $\phi(\lambda_1f_1+\lambda_2f_2)=\lambda_1\phi(f_1)+
\lambda_2\phi(f_2)$ for all $\lambda_i\in\CC$, $f_i\in\DD(M)$
\item Field Equation: $\phi((\Dal_\gb+\mu)f)=0$ for all 
$f\in\DD(M)$. 
\item CCR's: $[\phi(f_1),\phi(f_2)]=i \Delta_\gb(f_1,f_2)\II$
for all $f_i\in\DD(M)$.
\end{list}
Relation (Q4) implements the quantisation of the theory,
and also ensures that $\AAA(M,\gb)$ obeys the causality 
axiom of Wightman theory (see, e.g.,~\cite{Haag}) namely, that
$\phi(f_1)$ and $\phi(f_2)$ should commute whenever the $\supp f_i$ are 
spacelike separated with respect to $C_\gb^+$. 

If $(M,\gb,C_\gb^+)$ is not globally hyperbolic this
construction fails owing to the absence of a global
advanced-minus-retarded fundamental bisolution. There are, of course,
$*$-algebras of smeared fields on $M$ obeying relations
(Q1), (Q2) and (Q3), but these relations are not sufficient to
specify a reasonable quantum field theory on $(M,\gb)$. In~\cite{Kay},
Kay suggested the following {\em F-locality condition} as a possible
substitute for relation~(Q4).

\begin{Def}[F-locality]
A $*$-algebra $\AAA$ of smeared fields on $(M,\gb,C_\gb^+)$
obeying (Q1), (Q2) and (Q3) is said to be {\em F-local} (with respect to 
$C_\gb^+$) if every point of $M$ has an intrinsically globally
hyperbolic neighbourhood $N$ such that 
\begin{equation}
[\phi(f_1),\phi(f_2)]=i\Delta_{\gb|_N}(f_1,f_2)\II
\label{eq:FLcon}
\end{equation}
whenever the
$\supp f_i$ are contained in $N$, and where $\Delta_{\gb|_N}$
is the advanced-minus-retarded fundamental bisolution on
$(N,\gb|_N,C_\gb^+|_N)$. A spacetime $(M,\gb,C_\gb^+)$ which admits an 
F-local algebra is
said to be {\em F-quantum compatible} with real linear 
scalar field theory. 
\end{Def}

F-locality thus requires that (Q4) should hold locally. Note that the
time orientation plays an important r\^ole in this definition as it
fixes the local form of commutators. Thus a spacetime $(M,\gb)$ which
admits two global time orientations might be
F-quantum compatible with respect to both, one or neither. (Of course,
when these orientations are just mutual time reversals of each other
then either both are F-quantum compatible, or neither is.)

One may consider further possible replacements for (Q4). For
our purposes, the weaker notion of a {\em locally causal 
algebra} will be useful.
\begin{Def}[Local causality]
A $*$-algebra $\AAA$ of smeared fields on $(M,\gb,C_\gb^+)$
obeying (Q1), (Q2) and (Q3) is said to be
{\em locally causal} (with respect to $C_\gb^+$) if $\AAA$ is
nonabelian and every
point of $M$ has an intrinsically globally hyperbolic neighbourhood $N$
such that
\begin{equation}
[\phi(f_1),\phi(f_2)]=0 
\label{eq:LCcon}
\end{equation} 
whenever $\supp f_1$ and
$\supp f_2$ are spacelike separated from one another in $N$
with respect to $C_\gb^+$. A spacetime $(M,\gb,C_\gb^+)$ which admits a
locally causal algebra is said to be compatible with local causality. 
\end{Def}

Local causality is just the requirement that Wightman causality
should hold in a local fashion. We require that the algebra be
nonabelian in order to exclude the abelian classical field algebra 
for Klein--Gordon theory. While we do not propose local causality as
a sufficient condition for a quantum field algebra on $(M,\gb,C_\gb^+)$,
it is however a reasonable necessary condition, strictly
weaker than F-locality. 

It will be useful to make analogous definitions of local causality and
F-locality for $P_\gb$-bisolutions on $M$. In this paper we will say
that a $P_\gb$-bisolution $\Gamma$ is F-local with respect to $C_\gb^+$
if every point has an intrinsically globally hyperbolic neighbourhood
$N$ such that $\Gamma$ agrees with $\Delta_{\gb|N}$ on $N\times N$. The
bisolution $\Gamma$ will be called locally causal if each point has an
intrinsically globally hyperbolic neighbourhood $N$ such that $\Gamma$
vanishes on pairs of test functions whose supports are spacelike
separated in $N$ with respect to $C_\gb^+$. Neighbourhoods such as $N$
will be called {\em neighbourhoods of F-locality} or {\em local
causality} as appropriate. Note that an F-local (locally causal)
bisolution must satisfy additional requirements (antisymmetry, reality,
and nontriviality) in order to be the commutator of an F-local (locally
causal) algebra.\footnote{In~\cite{FH} these additional requirements
were included in the definition of an F-local bisolution.}

Let us observe that neither F-locality nor local causality assert any
uniformity in the `size' of the neighbourhoods of F-locality or local
causality in which Eqs.~(\ref{eq:FLcon}) or~(\ref{eq:LCcon}) hold. 
In addition, the
commutator $[\phi(f_1),\phi(f_2)]$ is not assumed to be a scalar
multiple of the identity unless the supports of $f_1$ and $f_2$ are
sufficiently small and close together. Remarkably, we will see that both
these properties will hold (at least generically) on the 2-dimensional 
cylinder manifold. 

\sect{Timelike and Spacelike Cylinders}\label{sect:TSC}

\subsection{Definitions}\label{sect:tns}

Here, we define the timelike and spacelike cylinders to be studied in
later sections. The cylinder manifold $M$ is the quotient of $\RR^2$ by 
the translation $(t,z)\mapsto (t,z+2\pi)$, where $(t,z)$ are Cartesian
coordinates on $\RR^2$. These coordinates induce smooth vector fields 
$\tb$ and $\zb$ on $M$ by the formulae
\begin{equation}
(\tb(f))\circ q = \frac{\partial}{\partial t}(f\circ q)
\qquad {\rm and}\qquad
(\zb(f))\circ q = \frac{\partial}{\partial z}(f\circ q),
\end{equation}
for $f\in C^\infty(M)$, where $q:\RR^2\to M$ is the defining quotient
map. We also use $q$ to define `diamond'
neighbourhoods $N_\epsilon(p)$ of each point $p\in M$ by
\begin{equation}
N_\epsilon(q(t_0,z_0)) = 
\left\{q(t,z)\mid |t-t_0|+|z-z_0|<\epsilon\right\},
\label{eq:diamnhd}
\end{equation}
and also to define the translations $T_{(\tau,\zeta)}$ on $M$ by
$T_{(\tau,\zeta)}:q(t,z)\mapsto q(t+\tau,z+\zeta)$. The translation
acts on a function $f$ on $M$ by
\begin{equation}
\left((T_{(\tau,\zeta)}f)\circ q\right)(t,z) = (f\circ q)(t-\tau,z-\zeta).
\end{equation}

Let $\GG$ be the class of smooth, bounded Lorentzian metrics $\gb$ on $M$
with $\gb(\zb,\zb)$ everywhere negative and bounded away from zero. In
particular, $\GG$
contains the Minkowskian metric $\etb$ obeying $\etb(\tb,\tb) = 
-\etb(\zb,\zb) = 1$ and $\etb(\tb,\zb) = 0$ at all points. 
Each $\gb\in\GG$ is globally time orientable in two
ways (modulo time reversal) determined by the continuous
cone fields $p\mapsto C_{\gb,p}^{T+}$ and $p\mapsto C_{\gb,p}^{S+}$
defined by
\begin{equation}
C_{\gb,p}^{T+} = \{ \vb\in T_p(M)\mid 
\gb_p(\vb,\vb)\ge 0~{\rm and}~\eb_p(\vb,\tb)\ge0\}
\end{equation}
and
\begin{equation}
C_{\gb,p}^{S+} = \{ \vb\in T_p(M)\mid 
\gb_p(\vb,\vb)\le 0~{\rm and}~\eb_p(\vb,\zb)\ge 0\},
\end{equation}
where $\eb$ is the background Euclidean metric defined by 
$\eb(\tb,\tb) = \eb(\zb,\zb) = 1$ and $\eb(\tb,\zb) = 0$ at all points
of $M$. We will refer to the triple $(M,\gb,\Cgt)$ as a {\em timelike
cylinder} and to $(M,\gb,\Cgs)$ as a {\em spacelike cylinder}. In
particular, $(M,\etb,\Cet)$ and $(M,\etb,\Ces)$ are the Minkowskian 
timelike and spacelike cylinders studied in~\cite{Kay,FH}.\footnote{In
fact, the Minkowskian spacelike cylinder was defined in~\cite{Kay,FH} as
the quotient of 2-dimensional Minkowski space with coordinates $(t,z)$
by the timelike
translation $t\mapsto t+T$ for some $T>0$ with the inherited causal
structure. This is equivalent to our present definition under
interchange of $t$ and $z$ in the case $T=2\pi$.} It is
easy to show that all timelike cylinders are globally hyperbolic with
Cauchy surfaces $q(\{t\}\times\RR)$. The corresponding
advanced-minus-retarded solution $\Delta_\gb$ will play an
important r\^ole in our discussion of both timelike and spacelike
cylinders. All the spacelike cylinders $(M,\gb,\Cgs)$ are
nonglobally hyperbolic because they contain closed timelike
lines (e.g., the integral curves of $\zb$, which are timelike with
respect to $\Cgs$) but because they do not possess initially
globally hyperbolic regions they evade the KRW theorems~\cite{KRW}. 

Let us note that the Klein--Gordon operator $P_\gb=\Dal_\gb+\mu$ is
locally hyperbolic on both timelike and spacelike cylinders. However,
the interpretation of $\mu$ depends on the time orientation: on the 
timelike cylinder
$\mu$ corresponds to the particle mass squared; on the spacelike
cylinder $\mu$ corresponds to minus the particle mass squared. Our
results in this paper hold for either sign of $\mu$, but {\em not} for
$\mu=0$. In addition, we will assume that $\mu$ is not a negative
integer to avoid unncessary technicalities. 

Our results will apply to timelike and spacelike cylinders whose metrics 
agree with $\etb$ outside a nonempty compact subset of $M$ of the form
$J_\etb^+(\{p\})\cap J_\etb^-(\{p'\})$ for $p,p'\in M$ 
(using $\Cet$ to define the causal future and past). Accordingly, we
define $\GG_K$ be the set of metrics in $\GG$ which agree with $\etb$
outside $K$. We will also discuss subclasses of $\GG_K$ which exhibit
some degree of $z$-translational symmetry. 
Let $G$ stand for either $\ZZ_N$ ($N\ge 2$) or ${\rm SO}(2)$, 
where $\ZZ_N$ is realised as the subgroup of $\RR/(2\pi\ZZ)$ consisting
of (equivalence classes of) $2\pi r/N$, for $r=0,\ldots,N-1$, and 
${\rm SO}(2)$ is realised as $\SSS^1=\RR/(2\pi\ZZ)$. The group $G$ acts
on $M$ by the corresponding `translations in $z$' 
$T_{(0,\zeta)}$ for $\zeta\in G$, 
and we use $\GG_K^{(G)}$ to denote the set of metrics $\gb\in\GG_K$
which are $G$-invariant, namely those obeying
\begin{equation}
\gb= T_{(0,\zeta)} \gb,
\end{equation}
for all $\zeta\in G$. 

To conclude this section, we describe a topology on $\GG_K$ and its
subclasses. The topology is defined so that 
$\gb_n\to\gb$ in $\GG_K$ if and only if 
$\gb_n^{-1}\to \gb^{-1}$ in $\EE^2_0(M)$, the Fr\'echet
space of tensor fields of type $(2,0)$. That is, $\gb_n\to\gb$
if and only if the components of $\gb_n^{-1}$ converge to
those of $\gb^{-1}$ in all coordinate patches. The significance of this
topology is that one may show (using energy norm arguments such as those
in \S 7.4 of~\cite{HawEl}) that the map 
\begin{equation}
\gb \mapsto \Delta_\gb f
\end{equation}
is continuous from $\GG_K$ to $\EE(M)$ for each fixed $f\in\DD(M)$, and
hence that matrix elements of an appropriate scattering operator 
vary continuously with the metric (see Sections~\ref{sect:KGM} and
Appendix~\ref{appx:Sgen}).  

\subsection{The Klein--Gordon Equation on $M$}\label{sect:KGM}

We now describe various spaces of (bi)solutions
for the Klein--Gordon operator $P_\gb$ for $\gb\in\GG_K$, exploiting 
the global hyperbolicity of $(M,\gb,\Cgt)$ and the
corresponding advanced-minus-retarded fundamental solution $\Delta_\gb$.  
The three solution spaces considered here are the space of smooth solutions
\begin{equation}
\FF_\gb = \{u\in\EE(M)\mid P_\gb u=0\},
\end{equation}
the space of weak solutions
\begin{equation}
\WW_\gb = \{\varphi\in\DD'(M)\mid \varphi(P_\gb f) = 0~\forall f\in\DD(M)\},
\end{equation}
and the space of weak bisolutions
\begin{equation}
\WW_\gb^{(2)} = \{\Gamma\in\DD^{(2)}{}'(M)\mid
\Gamma(P_\gb f_1,f_2)=\Gamma(f_1,P_\gb f_2) = 0~\forall f_1,f_2\in\DD(M)\}.
\end{equation}
We endow these spaces with the relative topologies inherited from
$\EE(M)$, $\DD'(M)$ and $\DD^{(2)}{}'(M)$
respectively. Since $(M,\gb,\Cgt)$ has compact Cauchy surfaces, 
we have $\FF_\gb=\Ran\Delta_\gb$ and dually, $\WW_\gb=\Ran\Delta_\gb'$
and $\WW_\gb^{(2)} = \Ran(\Delta_\gb\otimes\Delta_\gb)'$. 

Each weak solution $\varphi\in\WW_\gb$ defines a continuous linear
functional $\langle\varphi;\cdot\rangle_\gb$ on $\FF_\gb$ so that
\begin{equation}
\langle\varphi;\Delta_\gb f\rangle_\gb = \varphi(f)
\label{eq:WactsonF}
\end{equation}
for all $f\in\DD(M)$, and the map
$\varphi\mapsto\langle\varphi;\cdot\rangle_\gb$ is in fact an
isomorphism of the topological vector spaces $\WW_\gb$ and $\FF_\gb'$
when the latter space is given its weak-$*$ topology. Similarly, the map
$\Gamma\mapsto \langle\Gamma;\cdot\rangle_\gb^{(2)}$ given by
\begin{equation}
\langle\Gamma;\Delta_\gb f_1\otimes \Delta_\gb f_2\rangle_\gb^{(2)} =
\Gamma(f_1\otimes f_2)
\end{equation}
for $f_i\in\DD(M)$ is an isomorphism of $\WW_\gb^{(2)}$ with
$(\FF_\gb\otimes \FF_\gb)'$. 

The maps $\iota_\gb$ and
$\iota_\gb^{(2)}=\iota_\gb\otimes\iota_\gb$ embed $\FF_\gb$ and
$\FF_\gb\otimes\FF_\gb$ continuously and densely in $\WW_\gb$ and
$\WW_\gb^{(2)}$. Defining $\sigma_\gb:\FF_\gb\times\FF_\gb\to\CC$ by
\begin{equation}
\sigma_\gb(u,v) = \langle\iota_\gb u;v\rangle_\gb
\end{equation}
we have $\sigma_\gb(\Delta_\gb f_1,\Delta_\gb f_2) =
\langle\iota_\gb\Delta_\gb f_1;\Delta_\gb f_2\rangle_\gb = 
\Delta_\gb(f_1,f_2)$. Thus  $\sigma_\gb$ is the usual 
symplectic form on $\FF_\gb$ and may be written in the familiar form
\begin{equation}
\sigma_\gb(u,v) = \int_{\Sigma_t} \sqrt{h}\left(
v\nb(u)-u\nb(v)\right)
\end{equation}
for $u,v\in\FF_\gb$, where $\nb$ is the unit vector field normal to 
$\Sigma_t =q(\{t\}\times[0,2\pi])$, and $\sqrt{h}=|\gb(\zb,\zb)|^{1/2}$  
is the density derived from the induced metric on $\Sigma_t$.

The Cauchy evolution from the `in' region $M^-$ to the `out' region $M^+$
may be formulated as a scattering problem, and this will provide a
useful viewpoint in Section~\ref{sect:lcbs}. If $\gb\in\GG_K$ we
define wave operators $\Omega_\gb^\pm:\FF_\etb\to\FF_\gb$ so that for
each $u\in\FF_\etb$, $\Omega_\gb^\pm u $ is the unique element of
$\FF_\gb$ agreeing with $u$ in $M^\mp$. These maps are topological
vector space isomorphisms with the symplectic property 
$\sigma_\gb(\Omega_\gb^\pm u_1,\Omega_\gb^\pm u_2)
=\sigma_\etb(u_1,u_2)$ for all $u_i\in\FF_\etb$. The {\em scattering
operator} $S_\gb=(\Omega_\gb^-)^{-1}\Omega_\gb^+$  
of $P_\gb$ relative to $P_\etb$ is therefore a symplectic topological  
isomorphism on $\FF_\etb$ with the property that $u^+ = S_\gb u^-$ if
and only if there exists $u\in\FF_\gb$ agreeing with $u^\pm$ in $M^\pm$.  

The matrix elements of the scattering operator may be 
given explicitly in terms of $\Delta_\gb$: if $f^\pm\in\DD(M^\pm)$ 
and $u=\Delta_\etb f^-$ and $v=\Delta_\etb f^+$, then
$\Omega_\gb^+ u = \Delta_\gb f^-$, $\Omega_\gb^-v = \Delta_\gb f^+$
and
\begin{equation}
\sigma_\etb(S_\gb u,v) = \sigma_\gb(\Omega_\gb^+u,\Omega_\gb^-v) = 
\sigma_\gb(\Delta_\gb f^-,\Delta_\gb f^+)
=
\Delta_\gb(f^-,f^+).
\label{eq:Sexp}
\end{equation}
The scattering operator $S_\gb$ on $\FF_\etb$ extends continuously to a
map (also denoted $S_\gb$) on the space of weak solutions $\WW_\etb$.
Similarly, we define a scattering operator $S_\gb^{(2)}$ on
$\WW_\etb^{(2)}$ by extending the map $S_\gb\otimes S_\gb$ on 
$\FF_\etb\otimes\FF_\etb$. Thus, $\Gamma^+=S^{(2)}_\gb \Gamma^-$ if and 
only if there is a unique weak bisolution $\Gamma$ for $P_\gb$ agreeing
with $\Gamma^-$ on $M^-\times M^-$ and with $\Gamma^+$ on $M^+\times
M^+$. 

To conclude this section, we make the above discussion more
explicit in the particular case of the flat metric $\etb$ and
where $\mu$ is neither zero nor a negative integer. Here, for any
$f\in\DD(M)$, $\Delta_\etb f$ may be expressed as a
$\EE(M)$-convergent series
\begin{equation}
\Delta_\etb f = \sum_{\stack{n\in\ZZ}{\epsilon=\pm}}
i\epsilon \xi_{n\epsilon}(f)\xi_{-n-\epsilon}
\label{eq:Dact}
\end{equation}
where the functions $\xi_{n\epsilon}\in\EE(M)$ are given by
\begin{equation}
(\xi_{n\epsilon}\circ q)(t,z) = 
\frac{e^{-i\epsilon \omega_n t+ inz}}{(4\pi\omega_n)^{1/2}}
\end{equation}
with 
\begin{equation}
\omega_n^{1/2} =\left\{\begin{array}{cl} |n^2+\mu|^{1/4} & n^2>-\mu
\\ e^{i\pi/4}|n^2+\mu|^{1/4} & n^2<-\mu. \end{array}\right.
\end{equation}
In Eq.~(\ref{eq:Dact}), and in what follows, we have denoted the
distribution $\iota_\etb\xi_{n\epsilon}$ by $\xi_{n\epsilon}$ to
simplify the notation. Thus $\xi_{n\epsilon}(f)$ is shorthand for  
$(\iota_\etb\xi_{n\epsilon})(f)$. 

To prove~(\ref{eq:Dact}) we employ the ($\DD(M)$-convergent)
Fourier expansion  
$f=\sum_{n\in\ZZ} f_n$, where $(f_n\circ q)(t,z)=e^{-inz}\alpha_n(t)$ 
with $\alpha_n\in\CoinX{\RR}$ and observe that 
\begin{equation}
\left((\Delta_\etb^{\pm}f_n)\circ q\right)(t,z) =
e^{-inz}\int_{\mp\infty}^t dt'\,
\alpha_n(t')\frac{\sin \omega_n(t-t')}{\omega_n}.
\end{equation}
Hence $\Delta_\etb f_n =\sum_{\epsilon=\pm} 
i\epsilon \xi_{n\epsilon}(f_n)\xi_{-n\,-\epsilon}$ from
which~(\ref{eq:Dact}) follows.

Equation~(\ref{eq:Dact}) shows that the $\xi_{n\epsilon}$ form a basis 
for $\FF_\etb$ and also allows us to give analogous expansions for general
elements of $\WW_\etb$ and $\WW_\etb^{(2)}$. 
For $\varphi\in\WW_\etb$ and $f\in\DD(M)$ we have
\begin{equation}
\varphi(f) = \langle \varphi,\Delta_\etb f\rangle_\etb = 
\sum_{\stack{n\in\ZZ}{\epsilon=\pm}}
i\epsilon\langle\varphi,\xi_{-n-\epsilon}\rangle_\etb\xi_{n\epsilon}(f)
\end{equation} 
by Eq.~(\ref{eq:WactsonF})
so $\varphi$ may be expressed as the $\WW_\etb$-convergent series
\begin{equation}
\varphi= \sum_{\stack{n\in\ZZ}{\epsilon=\pm}}
i\epsilon\langle\varphi,\xi_{-n-\epsilon}\rangle_\etb 
\,\xi_{n\epsilon},
\label{eq:Wrep}
\end{equation}
in which the coefficients 
$\langle\varphi;\xi_{-n-\epsilon}\rangle_\etb$ grow no faster than
polynomially in $n$ owing to the local regularity of
$\varphi$. Applying~(\ref{eq:Wrep}) to the particular case
$\varphi=\xi_{n'\epsilon'}$, we also obtain the formula
\begin{equation}
\sigma_\etb(\xi_{n'\epsilon'},\xi_{-n\,-\epsilon})=
\langle 
\xi_{n'\epsilon'},\xi_{-n\,-\epsilon}\rangle_\etb = -i\epsilon
\delta_{nn'}\delta_{\epsilon\epsilon'} 
\label{eq:xiorth}
\end{equation}
for all $n,n'$, $\epsilon,\epsilon'$. We will make use of this 
formula in Section~\ref{sect:lcbs}. 

In an exactly analogous fashion, any $\Gamma\in\WW_\etb^{(2)}$ may be
expanded as
\begin{equation}
\Gamma=\sum_{\stack{n,n'\in\ZZ}{\epsilon,\epsilon'=\pm}}
-\epsilon\epsilon'\langle\Gamma;\xi_{-n-\epsilon}\otimes\xi_{-n'-\epsilon'}
\rangle_\etb^{(2)} \xi_{n\epsilon}\otimes\xi_{n'\epsilon'},
\label{eq:W2rep}
\end{equation}
with the series converging in $\WW_\etb^{(2)}$ and coefficients
$\langle\Gamma;\xi_{-n-\epsilon}\otimes\xi_{-n'-\epsilon'}
\rangle_\etb^{(2)}$ growing no faster than polynomially in $n,n'$. 
In particular,
\begin{equation}
\Delta_\etb = \sum_{\stack{n\in\ZZ}{\epsilon=\pm}}
i\epsilon \xi_{n\epsilon}\otimes\xi_{-n-\epsilon}.
\end{equation}

\sect{Locally Causal Bisolutions on $M$}\label{sect:lcbs}

We now employ the formalism developed in Section~\ref{sect:KGM} to
classify locally causal bisolutions on the cylinder manifold. 
Our argument consists of two steps. First, consider a
bisolution $\Gamma\in\WW_\gb^{(2)}$ ($\gb\in\GG_K$) which is locally
causal with respect to $C_\gb^+$ (standing for either $\Cgt$ or $\Cgs$). 
We will show that the local causality of $\Gamma$ implies that the
bisolutions $\Gamma^\pm$ for $P_\etb$ with which $\Gamma$ agrees in
$M^\pm\times M^\pm$ are both translationally invariant, in the sense
that 
\begin{equation}
\Gamma^\pm(T_{(\tau,\zeta)}f_1\otimes
T_{(\tau,\zeta)}f_2) = \Gamma^\pm(f_1\otimes f_2)
\end{equation}
for all $(\tau,\zeta)\in\RR^2$. Second, noting that $\Gamma^+ =
S_\gb^{(2)}\Gamma^-$, we use the scattering formalism of
Section~\ref{sect:KGM} to investigate the class of bisolutions in
$\WW^{(2)}_\etb$ whose
translational invariance is preserved by $S_\gb^{(2)}$. For generic
$\gb\in\GG_K^{(G)}$ we will show how this class may be parametrised by
distributions on the symmetry group $G$. In particular, if $G$ is
trivial, the only possibility is for $\Gamma$ to be a scalar multiple of 
$\Delta_\gb$. 

We begin by showing that local causality of $\Gamma$ implies
translational invariance of $\Gamma^+$. The proof for $\Gamma^-$ is
identical. Choose $\alpha>\pi/2$ and $p_1\in M^+$ such that the set
\begin{equation}
H=\left\{ T_{(\tau,\zeta)}p_1\mid \tau\in
[-\alpha,\alpha],\quad\zeta\in[0,2\pi] \right\}
\end{equation}
is contained in $M^+$. Because $H$ is compact and $\Gamma^+$ agrees
with $\Gamma$ in $H\times H$ we deduce the existence of an $\epsilon>0$
such that for all $p\in H$ the diamond neighbourhood $N_\epsilon(p)$ is
a neighbourhood of local causality for $\Gamma$. Now pick any $p_2$
spacelike separated from $p_1$ in $N_\epsilon(p_1)$. Since the causal
structure of $M^+$ is translationally invariant (due to invariance of
$\etb$) $T_{(\tau,\zeta)}p_2$ is spacelike separated from
$T_{(\tau,\zeta)}p_1$ in the $N_\epsilon(T_{(\tau,\zeta)}p_1)$ for all
$\tau\in[-\alpha,\alpha]$, $\zeta\in[0,2\pi]$. Thus $\Gamma^+$ must
vanish in a neighbourhood of
\begin{equation}
\label{eq:surf}
S = \left\{(T_{(\tau,\zeta)}p_1,T_{(\tau,\zeta)}p_2)\mid 
\tau\in [-\pi/2,\pi/2],\quad\zeta\in[0,2\pi] \right\}
\end{equation}
in $M\times M$ because each $N_\epsilon(T_{(\tau,\zeta)}p_1)$ is a
neighbourhood of local causality for $\Gamma$. We now apply the
following result to conclude that $\Gamma^\pm$ are translationally
invariant. 

\begin{Thm} \label{Thm:uc}
If $\mu$ is neither zero nor a negative integer and
$\Gamma_0$ is a weak $P_\etb$-bisolution vanishing on an open
neighbourhood of a surface of the form~(\ref{eq:surf}), then $\Gamma_0$
is translationally invariant. 
\end{Thm}
Theorem~\ref{Thm:uc} is proved in~\cite{F} using methods drawn from
Beurling's theory of interpolation~\cite{Beu}. The result does {\em not}
hold if $\mu=0$ (the requirement that $\mu$ is not a negative integer is
included for technical convenience). Now translational invariance of a
bisolution $\Gamma_0\in\WW^{(2)}_\etb$ entails that in the expansion
\begin{equation}
\Gamma_0 = \sum_{\stack{n,n'\in\ZZ}{\epsilon,\epsilon'=\pm}}
\gamma^{\epsilon\epsilon'}_{nn'} \xi_{n\epsilon}\otimes\xi_{n'\epsilon'},
\end{equation}
all coefficients vanish except those on the anti-diagonal $n=-n'$,
$\epsilon =-\epsilon'$. Accordingly, the bisolutions $\Gamma^\pm$ may be
expanded in the form
\begin{equation}
\Gamma^- = \sum_{\stack{n\in\ZZ}{\epsilon=\pm}}
i\epsilon\gamma_{n\epsilon}
\xi_{n\epsilon}\otimes\xi_{-n\,-\epsilon}
\end{equation}
and
\begin{equation}
\Gamma^+ = \sum_{\stack{n\in\ZZ}{\epsilon=\pm}}
i\epsilon\widetilde{\gamma}_{n\epsilon}
\xi_{n\epsilon}\otimes\xi_{-n\,-\epsilon}
\label{eq:G2}
\end{equation}
with polynomially bounded coefficients $\gamma_{n\epsilon}$ and 
$\widetilde{\gamma}_{n\epsilon}$. But we also have
\begin{equation}
\Gamma^+=S^{(2)}_\gb\Gamma^-
=\sum_{\stack{n\in\ZZ}{\epsilon=\pm}}
i\epsilon\gamma_{n\epsilon}
(S_\gb \xi_{n\epsilon})\otimes(S_\gb\xi_{-n\,-\epsilon}).
\label{eq:G3}
\end{equation}
On the one hand, therefore,
\begin{eqnarray}
\langle \Gamma^+;S_\gb\xi_{-n'\,-\epsilon'},
\xi_{n''\epsilon''}\rangle
&=& \sum_{\stack{n\in\ZZ}{\epsilon=\pm}}
i\epsilon\widetilde{\gamma}_{n\epsilon}
\sigma_\etb(\xi_{n\epsilon},S_\gb\xi_{-n'\,-\epsilon'})
\sigma_\etb(\xi_{-n\,-\epsilon},\xi_{n''\epsilon''})
\nonumber \\
&=& \widetilde{\gamma}_{n''\epsilon''}
\sigma_\etb(S_\gb\xi_{-n'\,-\epsilon'}, \xi_{n''\epsilon''})
\end{eqnarray}
by Eqs.~(\ref{eq:G2}) and~(\ref{eq:xiorth}) and antisymmetry of
$\sigma_\gb$, whilst on the other hand
\begin{eqnarray}
\langle \Gamma^+;S_\gb\xi_{-n'\,-\epsilon'},
\xi_{n''\epsilon''}\rangle
&=& \sum_{\stack{n\in\ZZ}{\epsilon=\pm}}
i\epsilon\gamma_{n\epsilon}
\sigma_\etb( S_\gb\xi_{n\epsilon},S_\gb\xi_{-n'\,-\epsilon'})
\sigma_\etb(S_\gb\xi_{-n\,-\epsilon},\xi_{n''\epsilon''})
\nonumber \\
&=& \gamma_{n'\epsilon'}
\sigma_\etb( S_\gb\xi_{-n'\,-\epsilon'},\xi_{n''\epsilon''})
\end{eqnarray}
using Eqs.~(\ref{eq:G3}) and~(\ref{eq:xiorth}) and the
symplectic property of $S_\gb$. Thus
\begin{equation}
\widetilde{\gamma}_{n''\epsilon''}
\sigma_\etb(S_\gb\xi_{-n'\,-\epsilon'},\xi_{n''\epsilon''})
=
\sigma_\etb(S_\gb\xi_{-n'\,-\epsilon'}, \xi_{n''\epsilon''})
\gamma_{n'\epsilon'}
\label{eq:itwine}
\end{equation}
for all $n',\epsilon',n'',\epsilon''$. We now invoke the following
result, which will be proved in Appendix~\ref{appx:Sgen}.
\begin{Thm} \label{Thm:Sgen}
For generic $\gb\in\GG_K^{(G)}$, we have
$\sigma_\etb(S_\gb\xi_{-n-\epsilon},\xi_{n'\epsilon'})\not =0$
whenever $n=n'$ (if $G={\rm SO}(2)$) or
$n\equiv n' \pmod{N}$ (if $G=\ZZ_N$).\footnote{In the case $N=1$,
$n\equiv n' \pmod{N}$ for all $n,n'\in\ZZ$.}
\end{Thm}
Thus, if $\gb$ belongs to the generic class marked out by
Theorem~\ref{Thm:Sgen}, we deduce from Eq.~(\ref{eq:itwine}) that
$\gamma_{n\epsilon}=\widetilde{\gamma}_{n'\epsilon'}$ whenever
$n\equiv n'\pmod{N}$ (if $G=\ZZ_N$) or $n=n'$ (if $G={\rm SO}(2)$).
Thus $\gamma_{n+}=\gamma_{n-}=\widetilde{\gamma}_{n+}
=\widetilde{\gamma}_{n-}=\gamma_n$ for all $n\in\ZZ$ with
$\gamma_n=\gamma_{n'}$ whenever $n\equiv n'\pmod{N}$ if $G=\ZZ_N$.
Reassembling $\Gamma^\pm$ we have
\begin{equation}
\Gamma^+=\Gamma^- = 
\sum_{\stack{n\in\ZZ}{\epsilon=\pm}}
i\epsilon \gamma_n
\xi_{n\epsilon}\otimes\xi_{-n\,-\epsilon}.
\label{eq:Gplus}
\end{equation}

The final step in our argument here is to write $\Gamma^\pm$ in terms of
a distribution on $G$. Here, we define the test functions $\DD(G)$ to be
the set of all complex valued functions on $G$ if $G=\ZZ_N$, and use the
usual space of test functions $\DD(\SSS^1)$ if $G={\rm SO}(2)$.
Now for any $f_1,f_2\in\DD(M)$ the formula
\begin{equation}
(f_1\diamond_\gb f_2)(\zeta) = \Delta_\gb(T_{(0,\zeta)}f_1,f_2)
\label{eq:diadef}
\end{equation}
defines a function $f_1\diamond_\gb f_2:G\to\CC$ belonging to $\DD(G)$. 
In particular, if $f_i\in\DD(M^-)$ we have
\begin{equation}
\label{eq:512}
(f_1\diamond_\gb f_2)(\zeta) = \Delta_\etb(T_{(0,\zeta)}f_1,f_2)
=\sum_{\stack{n\in\ZZ}{\epsilon=\pm}} 
i\epsilon e^{in\zeta} \xi_{n\epsilon}(f_1)\xi_{-n\,-\epsilon}(f_2),
\end{equation}
since $\gb=\etb$ in $M^-$. We claim that there exists a unique
$\psi\in\DD'(G)$ obeying $\psi(e_n)=\gamma_n$ for all $n\in\ZZ$, where
$e_n(\zeta)=e^{in\zeta}$. This is evident if $G={\rm SO}(2)$ because the
$\gamma_n$ are polynomially bounded. In the case $G=\ZZ_N$, we note that 
$n\equiv n'\pmod{N}$ entails that $\gamma_n=\gamma_{n'}$ and also that
$e_n$ and $e_{n'}$ agree as functions on $G$, so the conditions 
$\psi(e_n)=\gamma_n$
constitute $N$ independent conditions on the $N$-dimensional space
$\DD'(G)$. Comparing with Eq.~(\ref{eq:Gplus}) we observe that  
\begin{equation}
\Gamma^-(f_1,f_2) = \psi(f_1\diamond_\gb f_2).
\end{equation}
Now the right hand side of this formula defines a global bisolution
$\Gamma_\psi$ for $P_\gb$ because $G$-invariance of $\gb$
entails that $h_1\diamond_\gb 
(P_\gb h_2)  = (P_\gb h_1)\diamond_\gb h_2=0$, the zero function on $G$,
for any $h_i\in\DD(M)$. Since $\Gamma$ agrees with
$\Gamma_\psi$ in $M^-\times M^-$, we have $\Gamma=\Gamma_\psi$
identically. 

We may summarise the discussion above in the following statement. 
\begin{Thm}\label{Thm:Gps}
For generic $\gb\in\GG_K^{(G)}$ a bisolution $\Gamma\in\WW_\gb^{(2)}$ is
locally causal with respect to $\Cgt$ or $\Cgs$ only if
$\Gamma=\Gamma_\psi$ for some $\psi\in\DD'(G)$, where
$\Gamma_\psi(f_1,f_2) = \psi(f_1 \diamond_\gb f_2)$.
\end{Thm}

In particular, consider the case in which $G$ is trivial,
$G=\{0\}$. Here $\DD'(G)$ is the 1-dimensional space of 
scalar multiples of $\delta_0$, the $\delta$-distribution at $0\in G$.
Now if $\psi=\lambda\delta_0$ then $\psi(f_1\diamond_\gb f_2) =
\lambda\Delta_\gb(f_1,f_2)$ by Eq.~(\ref{eq:diadef}), so 
Theorem~\ref{Thm:Gps} asserts that $\Gamma$ is locally causal with
respect to $\Cgt$ or $\Cgt$ only if $\Gamma$ is a scalar multiple of
$\Delta_\gb$. Furthermore, since $\Delta_\gb$ is not locally causal with
respect to $\Cgs$ we conclude that there are no nontrivial locally
causal bisolutions on generic spacelike cylinders with metrics in
$\GG_K$.

\sect{Quantum Field Theory on Cylinder Spacetimes}\label{sect:results}

In this section we will state and prove our results concerning locally
causal and F-local algebras on the timelike and spacelike cylinders with
metrics belonging to our classes $\GG_K$ or $\GG_K^{(G)}$ for $G=\ZZ_N$
or ${\rm SO(2)}$. The essential idea is to reduce questions concerning
locally causal and F-local algebras to analogous questions concerning
the corresponding bisolutions, which can then be answered using the
results of Section~\ref{sect:lcbs}. We will only address the case in
which the mass parameter $\mu$ in the Klein--Gordon operator is neither
zero nor a negative integer, i.e., $\mu\not\in\ZZ^-$. 

Let $\gb\in\GG_K$ and suppose that $\AAA$ is a locally causal $*$-algebra
of smeared fields on $(M,\gb,C_\gb^+)$ where $C_\gb^+$ stands for either
$\Cgt$ or $\Cgs$. If $\omega$ is any element of the dual $\AAA'$, we may
define a bilinear functional $\Gamma$ on $\DD(M)\times\DD(M)$ by
\begin{equation}
\Gamma(f_1,f_2) = \omega([\phi(f_1),\phi(f_2)]).
\end{equation}
The functional $\Gamma$ is a bidistribution owing to the
topology of $\AAA$; it is also a locally causal bisolution for $P_\gb$ by
the field equation~(Q3) and local causality of $\AAA$. If $\gb$ belongs to
the generic subset of $\GG_K^{(G)}$ demarcated by Theorem~\ref{Thm:Gps} we
may deduce that $\Gamma=\Gamma_{\psi_\omega}$ for some
$\psi_\omega\in\DD'(G)$, so 
\begin{equation}\label{eq:62}
\omega([\phi(f_1),\phi(f_2)])=i\psi_\omega(f_1\diamond_\gb f_2)
\end{equation} 
for all $f_i\in\DD(M)$. Since $\omega$ was arbitrary and $\AAA'$ separates
the points of
$\AAA$, the commutator $[\phi(f_1),\phi(f_2)]$ can therefore
depend on $f_1$ and $f_2$ only through the combination $f_1\diamond_\gb
f_2$, so
\begin{equation}
[\phi(f_1),\phi(f_2)] = i\Psi(f_1\diamond_\gb f_2)
\label{eq:PScomm}
\end{equation}
for some linear $\Psi:\DD(G)\to\AAA$. The key step is provided by the
next Lemma. 

\begin{Lem}\label{Lem:Psi}
$\Psi(\cdot) = \psi(\cdot)\II$ for some $\psi\in\DD'(G)$. 
\end{Lem}
Lemma~\ref{Lem:Psi} will be proved using a Jacobi identity argument at
the end of this section. It follows that for generic
$\gb\in\GG_K^{(G)}$ the commutator in 
any locally causal algebra on $(M,\gb,C_\gb^+)$
takes the form
\begin{equation}
[\phi(f_1),\phi(f_2)] = i\psi(f_1\diamond_\gb f_2)\II =
i\Gamma_\psi(f_1,f_2)\II 
\label{eq:pscomm}
\end{equation}
for some $\psi\in\DD'(G)$ depending on $\AAA$. In the simplest case,
where $G$ is trivial, we obtain the following. 
\begin{Thm}\label{Thm:LC1}
Suppose $\mu\not\in\ZZ^-$. For generic $\gb\in\GG_K$, we have:
(1)~$\AAA$ is locally causal on
$(M,\gb,\Cgt)$ if and only if $\AAA=\AAA_\lambda(M,\gb)$, the quotient
of $\Af(M)$ by (Q1--3) and
\begin{list}{(Q\arabic{enumii})${}_\lambda$}{\usecounter{enumii}}
\setcounter{enumii}{3}
\item CCR's: $[\phi(f_1),\phi(f_2)]=i\lambda\Delta_\gb(f_1,f_2)\II$  
for all $f_i\in\CoinX{M;\RR}$
\end{list}
for some $\lambda\in\RR\backslash\{0\}$. (2)~There is no locally causal
algebra on $(M,\gb,\Cgs)$. 
\end{Thm}
{\noindent\em Proof:} Assume that $\gb$ belongs to the generic class of
Theorem~\ref{Thm:Gps}. 

{\noindent(1)}~The commutator takes
the form~(\ref{eq:pscomm}) for some
$\psi\in\DD'(G)$. Since $G=\{0\}$, we have
$\Gamma_\psi=\lambda\Delta_\gb$ for some $\lambda\in\CC$ by the
discussion at the end of Section~\ref{sect:lcbs}, so
$[\phi(f_1),\phi(f_2)]=i\lambda\Delta_\gb(f_1,f_2)\II$ for all
$f_i\in\DD(M)$. We deduce that $\lambda$ is real due to the
antihermiticity of the commutator, and since $\AAA$ is nonabelian we
must have $\lambda\not=0$. Thus $\AAA$ satisfies
relations (Q1--3) and (Q4)${}_\lambda$ and must be a quotient of
$\AAA_\lambda(M,\gb)$. But $\AAA_\lambda(M,\gb)$ is simple (see, e.g.,
\S7.1 of~\cite{Baez}), so $\AAA=\AAA_\lambda(M,\gb)$. 

{\noindent (2)}~If a locally causal algebra existed, its commutator
would be proportional to $\Delta_\gb$. Since this is not locally causal
with respect to $\Cgs$, the proportionality constant would vanish and
the algebra would be abelian -- a contradiction. $\square$

\begin{Cor} \label{Cor:FL1}
Suppose $\mu\not\in\ZZ^-$. 
For generic $\gb\in\GG_K$, we have: (1)~The usual field algebra
$\AAA(M,\gb)$ is the unique F-local algebra on $(M,\gb,\Cgt)$. 
(2)~The spacelike cylinder $(M,\gb,\Cgs)$ is F-quantum incompatible.
\end{Cor}
{\noindent\em Proof:} (1)~Comparing (Q4)${}_\lambda$ with the F-locality
condition, we require $\lambda=1$ for F-locality. (2)~This is immediate
from Theorem~\ref{Thm:LC1} because any F-local algebra is necessarily
locally causal.

These (perhaps surprising) results show that F-locality and local
causality are much stronger conditions than might have been expected on
these spacetimes. For generic (globally hyperbolic) timelike cylinders,
one is essentially restricted to
the usual algebra even in the locally causal case, because
$\AAA_\lambda(M,\gb)$ is related to $\AAA(M,\gb)$ by rescaling Planck's
constant by a factor of $|\lambda|$ and a change of time orientation
from $\Cgt$ to $-\Cgt$ if $\lambda<0$. 

The second parts of Theorem~\ref{Thm:LC1} and
Corollary~\ref{Cor:FL1} should be set against the fact that the
Minkowskian spacelike cylinder $(M,\etb,\Ces)$ admits infinitely many
F-local algebras for massive (and massless) Klein--Gordon
theory~\cite{FH}. It is clear we cannot replace `generic' by `all' in
the statements of these results, which raises the issue as to
whether there is a substantial -- albeit nongeneric in $\GG_K$ --
class of metrics whose corresponding spacelike cylinders are F-quantum
compatible. To address this, we refine the above results to cover
generic $G$-invariant metrics belonging to $\GG_K^{(G)}$ for $G=\ZZ_N$
($N\ge 2$) or ${\rm SO(2)}$. These spaces are of course nongeneric
subsets of $\GG_K$. Because the commutator has already been
established in Eq.~(\ref{eq:pscomm}),
it remains to determine more precisely the class of $\psi\in\DD'(G)$ for
which $\Gamma_\psi$ can determine commutators in a locally causal or
F-local algebra. We begin by noting that $\psi$ must be
nontrivial because such algebras are nonabelian. Next, $\psi$ is
constrained by the algebraic relations
\begin{equation}
[\phi(f_1),\phi(f_2)]=-[\phi(f_2),\phi(f_1)] \qquad{\rm and}\qquad
[\phi(f_1),\phi(f_2)]^* = -[\phi(\overline{f_1}),\phi(\overline{f_2})]
\end{equation}
which follow from~(Q1) and
imply that $\Gamma_\psi(f_1,f_2)=-\Gamma_\psi(f_2,f_1)$ and 
$\overline{\Gamma_\psi(f_1,f_2)} =
\Gamma_\psi(\overline{f_1},\overline{f_2})$. In consequence, 
$\psi$ is {\em even}, i.e., $\psi(\widetilde{\gamma})=\psi(\gamma)$ for
$\gamma\in\DD(G)$, where $\widetilde{\gamma}(\zeta)=\gamma(-\zeta)$ for
all $\zeta\in G$, and {\em real}, i.e.,
$\overline{\psi(\gamma)}=\psi(\overline{\gamma})$.   

The remaining requirement is that $\Gamma_\psi$ should be
locally causal or F-local with respect to $C_\gb^+$. In the
case $G=\ZZ_N$, $\psi\in\DD'(G)$ may be regarded
as a vector $\psi=(\psi_0,\psi_1,\ldots,\psi_{N-1})\in\CC^N$
with action
\begin{equation}
\psi(\gamma) = \sum_{r=0}^{N-1} \psi_r \gamma(2\pi r/N)
\end{equation}
on a function $\gamma:\ZZ_N\to \CC$. [Recall that $\ZZ_N$ is realised as
the group of equivalence classes of $2\pi r/N$, $r=0,\ldots,N-1$.] Thus 
\begin{equation}
\Gamma_\psi(f_1,f_2) = \sum_{r=0}^{N-1} \psi_r 
\Delta_\gb(T_{(0,2\pi r/N)}f_1,f_2),
\label{eq:ZNGamm}
\end{equation}
and if the $f_i$ are supported within a common sufficiently small
neighbourhood only the $r=0$ term in Eq.~(\ref{eq:ZNGamm})
contributes yielding $\Gamma_\psi(f_1,f_2) =
\psi_0\Delta_\gb(f_1,f_2)$. Thus $\Gamma_\psi$ is automatically locally
causal with
respect to $\Cgt$, and is F-local with respect to $\Cgt$ if and only if
$\psi_0=1$. On the other hand, $\Gamma_\psi$ is locally causal with
respect to $\Cgs$ if and only if $\psi_0=0$ and can never be F-local with
respect to this time orientation. A more involved 
analysis for ${\rm SO(2)}$ (which we will omit) leads to the
conclusions that $\Gamma_\psi$ is locally causal with respect to $\Cgt$
if and only if condition
\begin{list}{}{}
\item[(T)] $\psi$ vanishes on $W\backslash\{0\}$ for some neighbourhood
$W$ of $0\in G$
\end{list}
holds; that $\Gamma_\psi$ is F-local with respect to $\Cgt$
if and only if condition
\begin{list}{}{}
\item[(T$'$)] $\psi$ agrees with $\delta_0$, the $\delta$-distribution at
$0\in G$, in some neighbourhood $W$ of $0\in G$ 
\end{list}
holds; and that $\Gamma_\psi$ is locally causal with respect to $\Cgs$
if and only if condition
\begin{list}{}{}
\item[(S)] $\psi$ vanishes on some neighbourhood
$W$ of $0\in G$
\end{list}
holds. In this last case,
$\Gamma_\psi$ vanishes on {\em all} pairs of test functions supported in
sufficiently small common neighbourhoods, so no bisolution of
the form $\Gamma_\psi$ can be F-local with respect to $\Cgs$.
We note that the conditions (T), (T$'$) and (S) coincide with the
conditions given on $\psi_0$ in the case $G=\ZZ_N$ if one endows $G$
with the discrete topology (so that $\{0\}$ is a neighbourhood of $0$).
In particular, condition~(T) is trivially satisfied. 

The above discussion leads immediately to the following conclusions. 
\begin{Thm} \label{Thm:LC2}
Suppose $\mu\not\in\ZZ^-$. For generic 
$\gb\in\GG_K^{(G)}$ we have: (1)~An algebra $\AAA$ of smeared fields  
is locally causal on the timelike cylinder $(M,\gb,\Cgt)$ if and only if 
the commutation relation
\begin{list}{(Q\arabic{enumii})${}_\psi$}{\usecounter{enumii}}
\setcounter{enumii}{3}
\item Modified CCR's: $[\phi(f_1),\phi(f_2)]=i\Gamma_\psi(f_1,f_2)\II$  
for all $f_i\in\CoinX{M;\RR}$
\end{list}
holds for some nontrivial, real, even $\psi\in\DD'(G)$ obeying
condition~(T) above. (2)~An algebra $\AAA$ of smeared fields  
is locally causal on the spacelike cylinder $(M,\gb,\Cgs)$ if and only
if the conditions of
part~(1) hold with condition~(T) replaced by condition~(S). 
\end{Thm}
\begin{Cor} \label{Cor:FL2}
Suppose $\mu\not\in\ZZ^-$. For generic 
$\gb\in\GG_K^{(G)}$ we have: (1)~An algebra $\AAA$ of smeared fields  
is F-local on the timelike cylinder $(M,\gb,\Cgt)$ if and only if 
the conditions of
part~(1) of Theorem~\ref{Thm:LC2} hold with condition~(T) replaced by 
condition~(T$\,'$). (2)~The spacelike cylinder $(M,\gb,\Cgs)$ is
F-quantum incompatible. 
\end{Cor}
In fact, the locally causal algebras described part~(2) of
Theorem~\ref{Thm:LC2} are
`locally abelian' (due to condition (S)) and are not serious
candidates for a quantum field algebra on the spacelike cylinder.

To be more quantitative concerning the range of F-local algebras
admitted by a generic $\ZZ_N$-invariant timelike cylinder, we note that
$\psi=(\psi_0,\ldots \psi_{N-1})\in\DD'(\ZZ_N)$ is real if and only if
each $\psi_r\in\RR$; even if and only if $\psi_r=\psi_{N-r}$ for
$r=1,\ldots, N-1$; and F-local if and only if $\psi_0=1$. Regarding
$\DD'(\ZZ_N)$ as a $2N$-dimensional real vector space, the F-local
elements lie on a $[\frac{1}{2}N]$-dimensional hyperplane, where $[x]$
denotes the integer part of $x\in\RR$. 

Corollary~\ref{Cor:FL2} shows that the F-quantum compatible Minkowskian 
spacelike
cylinder $(M,\etb,\Cgs)$ (see~\cite{FH}) represents an extremely
special case even within classes of metrics exhibiting a high degree of
symmetry. We know of no metric other than $\etb$ in $\GG_K$ whose
corresponding spacelike cylinder is F-quantum compatible with massive
Klein--Gordon theory, and conjecture
that no such metric exists. The F-quantum compatibility of  
$(M,\etb,\Cgs)$ appears to rely in an essential way on its
invariance under all spacetime translations. 

To summarise our results in this section, we have shown that local
causality is a highly restrictive condition on the 2-dimensional
cylinder manifold. Generically in $\GG_K$, with no symmetry assumed,
we have essentially no more
freedom than is afforded by the usual construction of quantum field
theory in curved spacetime: there is a 1-parameter family of locally
causal algebras $\AAA_\lambda(M,\gb)$ 
on the globally hyperbolic timelike cylinder of which
exactly one is F-local, and the nonglobally hyperbolic spacelike
cylinder fails to be compatible with local causality. By invoking
symmetry, some degree of freedom is obtained and we have classified the 
resulting locally causal and F-local algebras for generic timelike
cylinders arising from the classes $\GG_K^{(G)}$ ($G=\ZZ_N$ or 
${\rm SO}(2)$). 
However the spacelike cylinders remain F-quantum incompatible for
generic metrics in these classes, and although nontrivial symmetry does 
permit the existence of locally causal algebras on the spacelike
cylinder, these are not realistic quantum field algebras because their
local structure is that of the abelian classical theory. 

It remains to prove Lemma~\ref{Lem:Psi}.

{\noindent\em Proof of Lemma~\ref{Lem:Psi}:} From Eq.~(\ref{eq:512}) we
have 
\begin{equation}
(h_1\diamond_\gb h_2)(\zeta)
=\sum_{\stack{n\in\ZZ}{\epsilon=\pm}} 
i\epsilon e^{in\zeta} \xi_{n\epsilon}(h_1)\xi_{-n-\epsilon}(h_2)
\end{equation}
if $h_1,h_2\in\DD(M^-)$.
In particular, if $h_{n\epsilon}\in\DD(M^-)$ is chosen so that
$\Delta_\etb h_{n\epsilon} = \xi_{-n-\epsilon}$ then 
$(h_{n\epsilon}\diamond_\gb f)(\zeta)=e^{in\zeta}\xi_{-n-\epsilon}(f)$
and therefore
\begin{equation}
[\phi(h_{n\epsilon}),\phi(f)] = i\Psi(e_n) \xi_{-n-\epsilon}(f)
\end{equation}
for all $f\in\DD(M^-)$ where $e_n\in\DD(G)$ is the function
$e_n(\zeta)=e^{in\zeta}$. 

For any $p\in M^-$, pick $\epsilon>0$ so that $N_{2\epsilon}(p)$ is a
neighbourhood of local causality for $\AAA$. For any $f_1$ supported in
$N_\epsilon(p)$ we may find $f_2$ with support spacelike separated from
that of $f_1$ in $N_{2\epsilon}(p)$, and therefore obeying
$[\phi(f_1),\phi(f_2)]=0$. Applying the Jacobi identity to $f_1$, $f_2$ and
$h_{n\epsilon}$, we obtain
\begin{eqnarray}\label{eq:610}
0&=& [\phi(f_1),[\phi(f_2),\phi(h_{n\epsilon})]] 
+ [\phi(f_2),[\phi(h_{n\epsilon}),\phi(f_1)]]
+ [\phi(h_{n\epsilon}),[\phi(f_1),\phi(f_2)]]
\nonumber \\
&=& -i[\phi(f_1),\Psi(e_n)] \xi_{-n-\epsilon}(f_2) 
+i[\phi(f_2),\Psi(e_n)] \xi_{-n-\epsilon}(f_1) .
\end{eqnarray}
Eliminating $[\phi(f_2),\Psi(e_n)]$ from these two equations
($\epsilon=\pm$) we have
\begin{equation}
[\phi(f_1),\Psi(e_n)] \left(\xi_{-n-}(f_2)\xi_{-n +}(f_1) -
\xi_{-n+}(f_2)\xi_{-n-}(f_1)\right) =0.
\end{equation}
Assuming that at least one of $\xi_{-n \pm}(f_1)$ is nonzero, we deduce
that $\phi(f_1)$ commutes with $\Psi(e_n)$ by varying $f_2$. If
$\xi_{-n -}(f_1)=\xi_{-n +}(f_1)=0$, then Eq.~(\ref{eq:610}) becomes
\begin{equation}
[\phi(f_1),\Psi(e_n)] \xi_{-n\pm}(f_2)=0
\end{equation}
and we obtain the same conclusion by varying $f_2$ again.
Since $f_1$ and $p$ were arbitrary we conclude (using a partition of
unity and property (Q2)) that this is true for any $f_1\in\DD(M^-)$ and 
therefore for any $f_1\in\DD(M)$ using properties of the Cauchy problem
and the field equation (Q3). Thus each $\Psi(e_n)$ ($n\in\ZZ$) commutes
with all the generators $\phi(f)$ of $\AAA$ and is therefore a scalar
multiple of the identity. A simple Fourier argument, exploiting
continuity of $[\phi(f_1),\phi(f_2)]$ in the $f_i$, now shows that all
commutators are scalar multiples of the identity, from which we
obtain $\Psi(\cdot)=\psi(\cdot)\II$ for some $\psi\in\DD'(G)$,
completing the proof.  
$\square$ 

\sect{Conclusion}\label{sect:concl}

In this paper we have given a full and rigorous treatment of F-local and
locally causal algebras for massive Klein--Gordon equation
on 2-dimensional cylinder spacetimes whose
metrics deviate from $\etb$ within a compact region. We have
seen that the only F-local algebra admitted by a generic timelike cylinder
is in fact the usual field algebra, and that the locally causal algebras
form a 1-parameter family related to the usual algebra by rescaling
Planck's constant and/or a reversal of time orientation. Generic
spacelike cylinders fail to be compatible with both F-locality and local
causality. A full discussion of these results, as well as others
concerning 4-dimensional spacelike cylinders will appear in~\cite{FHK}
(see also~\cite{KayR}); here we will make only a few brief remarks. 

Firstly, the content of our results is that the F-local and
locally causal theories coincide generically with the conventional
globally hyperbolic theory on  cylinder spacetimes. Thus any
algebraic framework for quantum
field theory on the spacelike cylinders must violate F-locality and even 
local causality: causality violation on a cosmic scale would in
principle be detectable by local measurements. 

Secondly, although our results give strong reason to believe that 
no spacelike cylinders other than the 
Minkowskian spacelike cylinder $(M,\etb,\Ces)$ are F-quantum
compatible with massive
Klein--Gordon theory, we have not completely
resolved this point. It may be that a more detailed scattering analysis
would shed further light on this issue. 

The F-quantum compatibility of $(M,\etb,\Ces)$ and the generic F-quantum
incompatibility of more general spacelike cylinders with massive
Klein--Gordon theory shows that F-quantum compatibility can be quite a
delicate property. Furthermore, neither the massless
2-dimensional case, nor the situation in 4-dimensions is fully
resolved~\cite{FHK,KayR}. It is clear, for example, that the massless
F-quantum compatibility of $(M,\etb,\Ces)$ is stable under general
conformal perturbations of the metric. 
A more detailed discussion of the current status of
F-locality will appear in~\cite{FHK}.

{\noindent\em Acknowledgments:} As mentioned in the text, this paper is 
an outgrowth of work with Atsushi Higuchi and Bernard Kay~\cite{FHK}. 
It is a pleasure to thank both of them for many discussions on this
subject, and particularly Atsushi for first suggesting the use of
scattering methods in this problem. I am also grateful to
Simon Eveson, for many conversations concerning topological vector
spaces. The bulk of this work was
supported by EPSRC Grant No.~GR/K 29937 to the University of York. 

\appendix
\sect{Proof of Theorem~\ref{Thm:Sgen}}
\label{appx:Sgen}

{\noindent\bf Theorem~\ref{Thm:Sgen}\ }{\em
For generic $\gb\in\GG_K^{(G)}$, we have
$\sigma_\etb(S_\gb\xi_{-n-\epsilon},\xi_{n'\epsilon'})\not =0$
whenever $n=n'$ (if $G={\rm SO}(2)$) or
$n\equiv n' \pmod{N}$ (if $G=\ZZ_N$).}

{\noindent\em Proof:} 
Recall that a subset of a topological space is said to be generic if
it contains a countable intersection of open dense sets. Thus it is
enough to show that for each choice of $n,n'$, $\epsilon,\epsilon'$
with $n\equiv n' \pmod{N}$ (if $G=\ZZ_N$) or $n=n'$ (if
$G={\rm SO}(2)$),
the set 
\begin{equation}
{\cal N}=\{\gb\in\GG_K^{(G)}\mid 
\sigma_\etb( S_\gb\xi_{-n-\epsilon},\xi_{n'\epsilon'})\not =0\}
\end{equation} 
is open and dense in $\GG_K^{(G)}$. The set ${\cal N}$ is open because
$\gb\mapsto\sigma_\etb(S_\gb u,v)$ is continuous for fixed $u,v\in
\FF_\etb$ by Eq.~(\ref{eq:Sexp}) and the fact that --
as mentioned in Section~\ref{sect:tns} -- the
map $\gb\mapsto \Delta_\gb f$ is continuous from $\GG_K$ to $\EE(M)$ for 
each fixed $f\in\DD(M)$. To establish density of ${\cal N}$, we 
fix $\gb\in\GG_K^{(G)}$ and show that it can be approximated from 
within ${\cal N}$. We will use the following Lemma, which is proved at
the end of this appendix. 

\begin{Lem} \label{lem:inter}
Let $u^+=\Omega_\gb^+\xi_{-n-\epsilon}$ and
$v^-=\Omega_\gb^-\xi_{n'\epsilon'}$. Then
\begin{equation}
\supp u^+\cap \supp v^- \cap K
\end{equation}
has nonempty interior. 
\end{Lem}

Now because $\gb$ is $G$-invariant, we have
\begin{equation}
T_{(0,\zeta)}u^+ = e^{in\zeta}u^+ \qquad {\rm and} \qquad
T_{(0,\zeta)}v^- = e^{-in'\zeta}v^-
\end{equation}
for $\zeta\in G$. Our conditions on $n$ and $n'$ entail that the 
product $u^+v^-$ is $G$-invariant, so
Lemma~\ref{lem:inter} allows us to
pick a $G$-invariant $V\in\CoinX{{\rm int}\,K;\RR}$ for which
\begin{equation}
\int_M g^{1/2} u^+ V v^- \not = 0 ,
\label{eq:Vchoi}
\end{equation}
and which defines a 1-parameter
family $\{\gb_\lambda\mid\lambda\in\RR\}$ 
of conformal perturbations of $\gb$ by
\begin{equation}
\gb_\lambda = \left(1+\frac{\lambda V}{\mu}\right) \gb.
\end{equation}
One may check that $\gb_\lambda\in\GG_K^{(G)}$ for all sufficiently
small $\lambda$, and $\gb_\lambda\to\gb$ as $\lambda\to 0$. Since 
\begin{equation}
P_{\gb_\lambda}=\left(1+\frac{\lambda
V}{\mu}\right)^{-1}\left(P_\gb+\lambda V\right)
\end{equation}
the corresponding scattering operators $S_{\gb_\lambda}$ are equal to
those for $P_\gb+\lambda V$ relative to $P_\gb$ for sufficiently small 
$\lambda$. Applying a Born expansion, we deduce that the function
$\lambda\mapsto
\sigma_\etb(S_{\gb_\lambda}\xi_{-n-\epsilon},\xi_{n'\epsilon'})$ is 
differentiable at $\lambda=0$ with derivative
\begin{equation}
\left.\frac{d}{d\lambda}
\sigma_\etb(S_{\gb_\lambda}\xi_{-n-\epsilon},\xi_{n'\epsilon'})
\right|_{\lambda=0} = -\int_M g^{1/2} u^+ V v^-,
\end{equation}
which is nonzero by Eq.~(\ref{eq:Vchoi}). We infer that 
$\sigma_\etb(S_{\gb_\lambda}\xi_{-n-\epsilon},
\xi_{n'\epsilon'})\not=0$ and therefore $\gb_\lambda\in{\cal N}$
for all sufficiently small $\lambda\not=0$. 
Consequently, every neighbourhood of $\gb$ in $\GG_K$ has nontrivial 
intersection with ${\cal N}$, and we conclude that ${\cal N}$ is 
dense as required. $\square$

{\noindent\em Proof of Lemma~\ref{lem:inter}:} 
We will use the following facts:
\begin{enumerate}
\item Since $v^-=\Omega_\gb^-\xi_{n'\epsilon'}$ agrees with
$\xi_{n'\epsilon'}$ in $M^+=M\backslash J_\etb^-(K)$ and is therefore
bounded away from zero there, $v^-$ is nonvanishing in some neighbourhood
of any point on $\partial J_\etb^-(K)$ by continuity. 
Similarly, $u^+$ is nonvanishing in some neighbourhood of any
point on $\partial J_\etb^+(K)$. 
\item $v^-$ cannot vanish identically in $M^-$
(otherwise it would vanish everywhere in $M$). As $M^-=M\backslash
J_\etb^+(K)$ belongs to the past domain of dependence of
$\partial J_\etb^+(K)$, $v^-$ is nonvanishing in some neighbourhood
of a point on $\partial J_\etb^+(K)$. Similarly, $u^+$ is
nonvanishing in some neighbourhood of a point on $\partial J_\etb^-(K)$.

\end{enumerate}

There are now two cases, depending on the geometry of $K=
J_\etb^+(\{p\})\cap J_\etb^-(\{p'\})$. Choose points $(t_p,z_p)$,
$(t_{p'},z_{p'})$ in the covering space of $M$ such that
$p=q(t_p,z_p)$, $p'=q(t_{p'},z_{p'})$ and $|z_{p'}-z_{p}|\le\pi$.
As $K$ is nonempty and $p\not=p'$, we have $t_{p'}>t_{p}$. 
Then we have
\begin{eqnarray}
J_\etb^+(\{p\}) &=& \{q(t,z)\mid t-t_p\ge |z-z_p|\} \nonumber \\
\partial J_\etb^+(\{p\}) &=& 
\{q(t_p+\chi,z_p\pm\chi)\mid 0\le\chi\le\pi\}
\end{eqnarray}
and
\begin{eqnarray}
J_\etb^-(\{p'\}) &=& \{q(t,z)\mid t-t_{p'}\ge |z-z_{p'}|\} \nonumber \\
\partial J_\etb^-(\{p'\}) &=& 
\{q(t_{p'}-\chi,z_{p'}\pm\chi)\mid 0\le\chi\le\pi\}.
\end{eqnarray}

{\noindent\em Case (i):} If $t_{p'}-t_p+|z_{p'}-z_{p}|>2\pi$
then $\partial J_\etb^+(K)$ and  $\partial J_\etb^-(K)$ do not intersect
and $\partial K = \partial J_\etb^+(K)\cup \partial J_\etb^-(K)$. 
Using the second fact
above, we pick $p''\in\partial J_\etb^+(K)$ such that $v^-$ is
nonvanishing in a neighbourhood of $p''$; by the first fact,
$u^+$ must also be nonvanishing in a neighbourhood of $p''$.
Since $p''\in\partial K$, 
the intersection of these neighbourhoods intersects $K$,
and $\supp u^+\cap \supp v^-\cap K$ has nonempty
interior.

{\noindent\em Case (ii):} If $t_{p'}-t_p+|z_{p'}-z_{p}|\le 2\pi$
then $\partial J_\etb^+(K)\cap \partial J_\etb^-(K)\cap \partial K$ 
consists of the two points
\begin{equation}
q\left(\frac{t_{p'}+t_p\pm(z_{p'}-z_p)}{2},
\frac{z_{p'}+z_p\pm(t_{p'}-t_p)}{2}\right)
\end{equation}
each of which has a neighbourhood in which both $u^+$ and $v^-$
are nonvanishing by the first fact above. Hence 
$\supp u^+\cap \supp v^-\cap K$ 
has nonempty interior as required.  $\square$


\begin{thebibliography}{ZZ}

\setlength{\itemsep}{0pt}\setlength{\parsep}{0pt}



\bibitem{Kay}    B.S. Kay, 
                 ``The principle of locality and quantum field theory
                 on (non globally hyperbolic) curved spacetimes''
                 Rev. Math. Phys. Special Issue (1992) 167--195 

\bibitem{Yurt}   U. Yurtsever,
                 ``Algebraic approach to quantum field theory on 
                 non-globally-hyperbolic spacetimes''
                 Class. Quantum Grav. {\bf 11} (1994) 999--1012

\bibitem{FPS}    J.L. Friedman, N.J. Papastamatiou and J.Z. Simon,
                 ``Failure of unitarity for interacting fields on
                 spacetimes with closed timelike curves''
                 Phys. Rev. D {\bf 46} (1992) 4456--4469

\bibitem{HawkSS} S.W. Hawking,
                 ``Quantum coherence and closed timelike curves''
                 Phys. Rev. D {\bf 52} (1995) 5681--5686

\bibitem{FHW}    C.J. Fewster, A. Higuchi and C.G. Wells,
                 ``Classical and quantum initial value problems for
                 models of chronology violation''
                 Phys. Rev. D {\bf 54} (1996) 3806--3825

\bibitem{YCB}    Y. Choquet-Bruhat, 
                 ``Hyperbolic partial differential equations on a
                 manifold'', in {\em Battelle Rencontres} 
                 C. De Witt and J. Wheeler (eds)
                 (Benjamin, New York, 1977)

\bibitem{Dim}    J. Dimock, 
                 ``Algebras of local observables on a manifold''
                 Commun. Math. Phys. {\bf 77} (1980) 219--228

\bibitem{HawkCP} S.W. Hawking,
                 ``The chronology protection conjecture''
                 Phys. Rev. D {\bf 46} (1992) 603--611

\bibitem{KRW}    B.S. Kay, M.J. Radzikowski and R.M. Wald,
                 ``Quantum field theory on spacetimes with a 
                 compactly generated cauchy horizon''
                 Commun. Math. Phys. {\bf 183} (1997) 533--556

\bibitem{CraKa1} C.R. Cramer and B.S. Kay,
                 ``Stress-energy must be singular on the Misner space
                 horizon even for automorphic fields''
                 Class. Quantum Grav. {\bf 13} (1996) L143--L149

\bibitem{CraKa2} C.R. Cramer and B.S. Kay,
                 ``Thermal and two-particle stress-energy must be ill   
                 defined on the two-dimensional Misner space chronology
                 horizon''
                 Phys. Rev. D {\bf 57} (1998) 1052--1056

\bibitem{FH}     C.J. Fewster and A. Higuchi,
                 ``Quantum field theory on certain non-globally
                 hyperbolic spacetimes''
                 Class. Quantum Grav. {\bf 13} (1996) 51--61

\bibitem{FHK}    C.J. Fewster, A. Higuchi and B.S. Kay,
                 ``How generic is F-locality? Examples and counterexamples''
                 In preparation

\bibitem{GChoqu} G. Choquet,
                 {\em Lectures on analysis, Volume 1: Integration and
                 topological vector spaces},
                 (Benjamin, New York, 1969)

\bibitem{F}      C.J. Fewster,
                 ``A unique continuation result for Klein--Gordon
                 bisolutions on a 2-dimensional cylinder''
                 {\tt gr-qc/9804011}

\bibitem{Dieud3} J. Dieudonn\'e,
                 {\em Treatise on analysis, Vol.~III},
                 (Academic Press, New York, 1972)

\bibitem{RSi}    M. Reed and B. Simon,
                 {\em Methods of modern mathematical physics Vol.~I:
                 Functional analysis}  
                 (Academic Press, New York, 1972)

\bibitem{Horm}   L. H\"ormander,
                 {\em The analysis of linear partial differential
                 operators, Vol.~I}
                 (Springer, Berlin, 1990) 

\bibitem{Schaef} H.H. Schaefer,
                 {\em Topological vector spaces},
                 (Macmillan, New York, 1966) 

\bibitem{Dieud7} J. Dieudonn\'e,
                 {\em Treatise on analysis, Vol.~VII},
                 (Academic Press, New York, 1988)

\bibitem{ONeill} B. O'Neill,
                 {\em Semi-Riemannian geometry},
                 (Academic Press, New York, 1983)

\bibitem{Dieck}  J. Dieckmann,
                 ``Cauchy surfaces in a globally hyperbolic space-time''
                 J. Math. Phys. {\bf 29} (1988) 578--579

\bibitem{Haag}   R. Haag,
                 {\em Local quantum physics: Fields, particles, algebras},
                 (Springer-Verlag, Berlin, 1992)                 

\bibitem{HawEl}  S.W. Hawking and G.F.R. Ellis,
                 {\em The large scale structure of space-time},
                 (Cambridge University Press, Cambridge, 1973)

\bibitem{Beu}    A. Beurling, 
                 ``Mittlag-Leffler Lectures on harmonic analysis (1977-1978)''
                 in {\em The collected works of Arne Beurling, Volume 2:
                 Harmonic analysis}, 
                 L. Carleson {\em et al.} (eds) 
                 (Birkh\"auser, Boston, 1989)

\bibitem{Baez}   J.C. Baez, I.E. Segal and Z. Zhou,
                 {\em Introduction to algebraic and constructive quantum
                 field theory},
                 (Princeton University Press, Princeton, 1992)

\bibitem{KayR}   B.S. Kay, 
                 ``Quantum fields in curved spacetime: non global
                 hyperbolicity and locality'' in {\em Operator algebras
                 and quantum field theory} 
                 S. Doplicher {\em et al.} (eds) 
                 (International Press)
                 {\tt gr-qc/9704075}

\end{thebibliography}
\end{document}